# Eye Care You: Voice Guidance Application Using Social Robot for Visually Impaired People

*Ting-An Lin* (林亭安), *Pei-Lin Tsai* (蔡姵琳), *Yi-An Chen* (陳羿安), *Feng-Yu Chen* (陳鳳妤),

and **Lyn Chao-ling Chen* (陳昭伶)

Department of Information Management,
Fu Jen Catholic University, New Taipei City, Taiwan
*E-mail: lynchen@ntu.edu.tw

**ABSTRACT**

In the study, the device of social robot was designed for visually impaired users, and along with a mobile application for provide functions to assist their lives. Both physical and mental conditions of visually impaired users are considered, and the mobile application provides functions: photo record, mood lift, greeting guest and today highlight. The application was designed for visually impaired users, and uses voice control to provide a friendly interface. Photo record function allows visually impaired users to capture image immediately when they encounter danger situations. Mood lift function accompanies visually impaired users by asking questions, playing music and reading articles. Greeting guest function answers to the visitors for the inconvenient physical condition of visually impaired users. In addition, today highlight function read news including weather forecast, daily horoscopes and daily reminder for visually impaired users. Multiple tools were adopted for developing the mobile application, and a website was developed for caregivers to check statues of visually impaired users and for marketing of the application.

## 1. INTRODUCTION

In the study, both physical condition and mental condition of the visually impaired people are considered. Home safety and social interaction of the visually impaired people are noticed in modern society, and the *Eye Care You* was designed for home care and social persuasion of visually impaired users (Figure 1 (a) and (b)). Normally, the visually impaired people have poor physical condition, for the lack of visual ability causes many inconveniences in their lives. Moreover, they also have the requirements of social activities. It reflects high index of loneliness of visually impaired elders, mental state of the visually impaired students, and quality of life of visually impaired in young ages, and the influences on the lives of young ages is greater than the elders [1]. From the data of socio-demographics, V-I characteristics, adverse life events, loneliness and life satisfaction, loneliness in the visually impaired people are higher in all ages than normal people, and it correlates with younger age, blindness, other disabilities, unemployment, bullying or abuse [1]. Therefore, the high loneliness score reduces the low level of life satisfaction. Hence, in the study, the *Eye Care You* with the guidance system and social interaction functions has been developed, for helping and improving visually impaired users in home scenario.

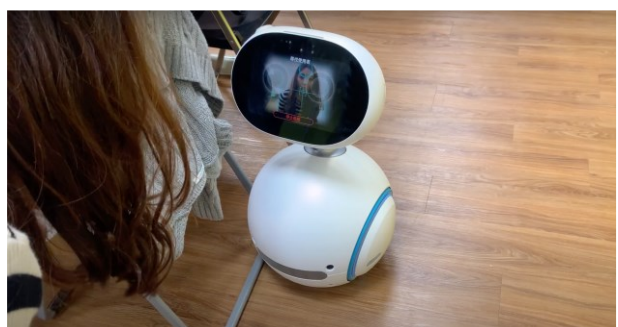

(a)

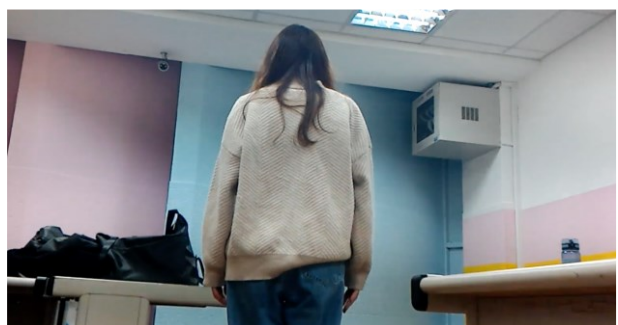

(b)

Fig. 1. Interaction between the user and the *Eye Care You*: (a) user identification and (b) user following.

**1.1. Physical condition of visually impaired people**

Physical condition of the visually impaired people has been discussed in the study, and one of the major problems is aging. From the study, it declared that the marriage state has the indirect effect of loneliness of visually impaired elders and mental dependence of their daily lives, including married, unmarried and the spouse who has passed away [2]. It also mentioned that they can get social support by expanding their moments for improving the loneliness situation [2]. Moreover, the study explained that the visually impaired elders between 55 to 84 years old has willingness to accept nursing care and prefer a double room, for the lack of support from children, the need of company from others and physical weakness [3].

**1.2. Mental condition of visually impaired people**

Mental condition of the visually impaired people has been discussed in the study. Mental health of the visually impaired people declines obviously than normal people, and it also reflects worse situation of patients with associated macular degeneration than the adults with glaucoma [4]. Moreover, the research focused on the mental health of visually impaired children using six scales of emotional stability, overall adjustment, autonomy, safety and insecurity and self-concept and IQ [5]. The emotional stability used MHB-VI as a screen test to examine visually impaired children with poor mental state, and Mental Health Battery was used to evaluate interventions effect of the program for reducing dropout rates and improving academic achievement [5]. A study interviewed visually impaired people of college students to discuss the interpersonal troubles since elementary school, and the problem had improved significantly after the sophomore year for adapting and experiencing three stage effects [6]. Another study explored the mental condition of people who are not total blind, and discussed the changes of their daily lives for helping them to adapting the society, such as the reconstruction of mental adaptation, the computer skills training, the overcoming mental fear before directional action training, the development of workplace action skills, and the visual experiences for selfcare [7].

**1.3. Quality of life of visually impaired people**

The QoL (Quality of Life) is a subjective index for revealing the type and duration of impairment and the participation in psychosocial rehabilitation. Generally, normal people have the better QoL indexes than the visually impaired people. There are differences in the QoL indexes among visually impaired people for the occurrence of visual impairment, and the QoL indexes in the congenital blindness group is better than the acquired blindness group [8]. The QoL indexes in the visually impaired people of working age adults (18 to 65 years old) reflect that the influences of disabilities in their daily lives are greater than the elders over 65 years old [9]. Moreover, the factors of personality, social support and the QoL index affect the visually impaired people for living a normal or near normal life [10]. Hence, both physical and mental conditions of visually impaired people are discussed in the study.

## 2. RRELATED WORKS

Currently, only few researches focused on the topic of social persuasion for visually impaired people, and that also brings out the value of the social robot in the study. In the design of hardware assistive device, a system implied connection and switch properties to the functions of the system for visually impaired users to turn on or turn off the functions quickly [11]. In addition, large icons in the interface design provided an intuitive experience for visually impaired users [12]. Social robot is also designed for visually impaired users to activate social cognitive mechanisms of users for the human-like appearance, thoughts and behaviors, comparing with the non-human technology interaction [13]. In the design of autonomous robot must consider the sociocultural environment of users to improve interaction between human and the robot, for understanding local sociocultural practices and social norms, interpreting and predicting user behavior [14]. For the properties of social robot, it is also helpful in health care to deliver the technology assist, interaction and response. Social Assistive Robots (SARs) are robotic platforms with audio, visual, and motor elements to interact with individuals for improving physical and mental health of users [15]. From the observation of human behavior, people usually turn around when they walk past others, and the robot simulates the behavior when it encounters users for acting natural and humanlike behavior [16]. Hence, social robot is helpful for visually impaired users for improving their daily lives in home scenario, both considering physical and mental aspects.

## 3. SYSTEM DESIGN

**3.1. System architecture**

System architecture of the *Eye Care You* showing in Figure 2, it consists of functions of the social robot and website development. ASUS Zenbo robot was used in the study for the extensibility of customized functions. The camera and microphone sensors of the Zenbo robot also enable vocal input and control of visually impaired users to record images in any situation. In addition, Zenbo App Builder was adopted for developing four main functions: photo interaction, mood lift, greeting guests and highlights of today, respectively (Figure 3). Website was designed for caregivers of visually impaired users. After registration an account, they can browse the record images according the date.

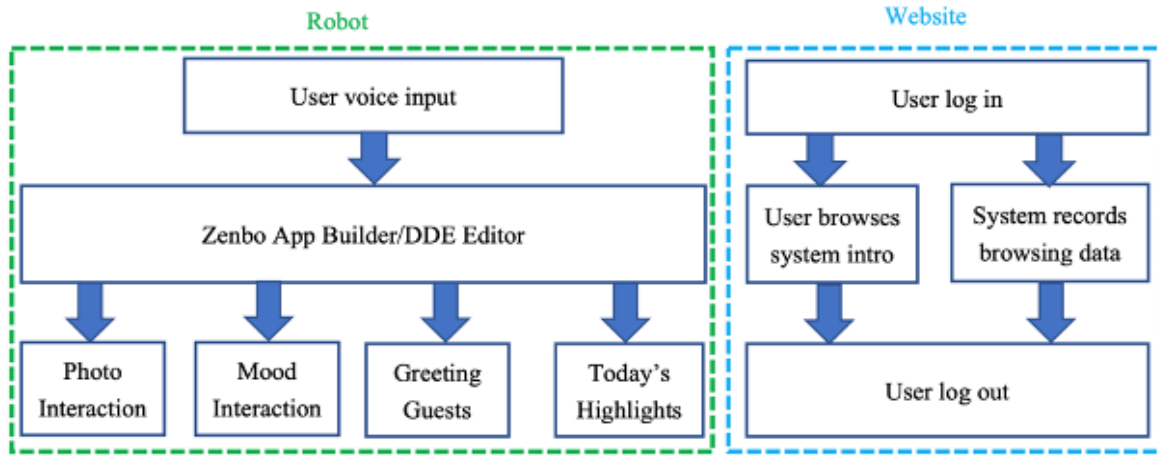

Fig. 2. System architecture of the *Eye Care You*.

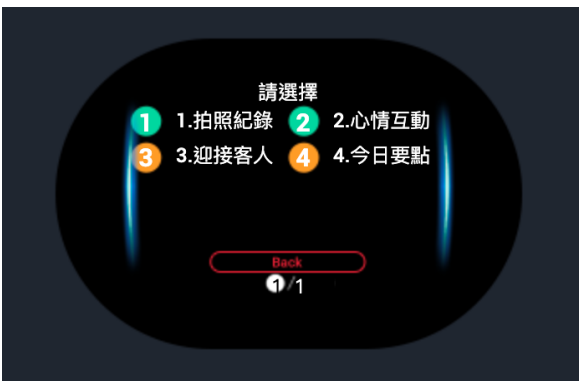

Fig. 3. Social interaction interface via voice control with four main functions: photo record, mood life, greeting guest and today highlight.

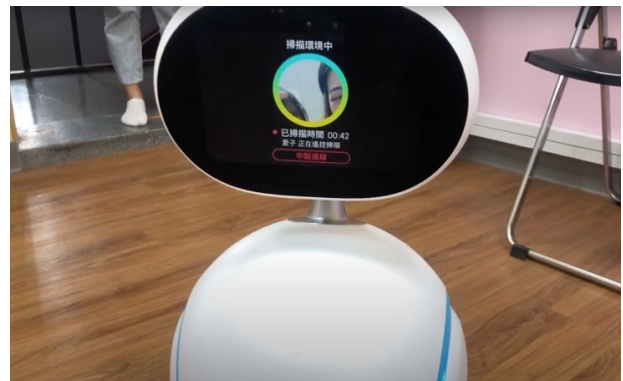

Fig. 4. Environment scan using a cellphone for remote control.

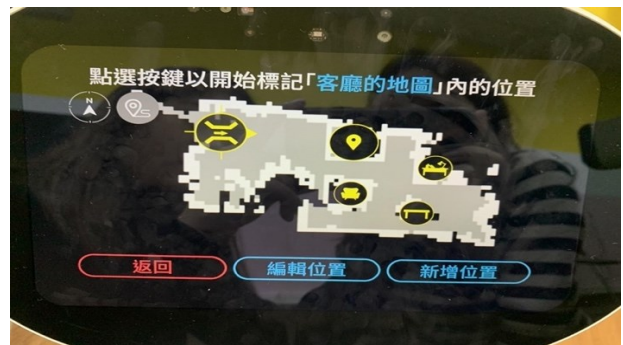

Fig. 5. Indoor map with marks of specific locations form users.

### 3.2. Social robot design

The market positioning of ASUS Zenbo robot is designed for using in home scenario, and the provided functions are also useful for visually impaired users at home. The ASUS Zenbo APP can create a map of home environment by scanning in 360-degree through remote control method of a cellphone (Figure 4), and it allows users to mark specific locations and obstacle areas on the indoor map (Figure 5). The environment setting of the map helps the Zenbo robot to move to precise positions. The ASUS Zenbo APP also allows caregivers to check the status of visually impaired users.

#### 3.2.1. Photo record, mood lift and greeting guest

Zenbo APP Builder was adopted in the study for designing the social interaction functions: photo record, mood lift and greeting guest. It is a graphical programming tool for facilitating system development. Based on the graphical editing model of Google Blockly, Zenbo APP Builder uses colorful blocks to write program, including action control, expressions, sensors and movements of the *Eye Care You*.

In photo record function, if visually impaired users encounter any danger or emergency situations, they can take photos immediately via voice control, or take photos for periodical record (Figure 6). Their caregivers can check the status of visually impaired users at home from the Google cloud album. In addition, the photos also help the caregivers to aware the potential risk and to remove obstacles in the home environment.

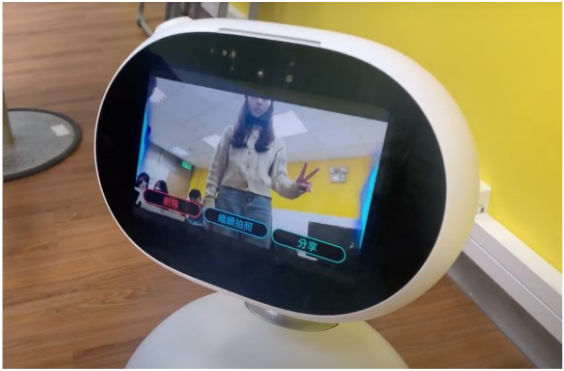

Fig. 6. Social interaction function: photo record.

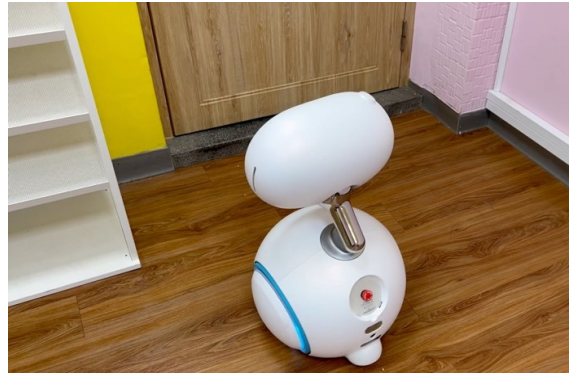

Fig. 8. Social interaction function: greeting guest.

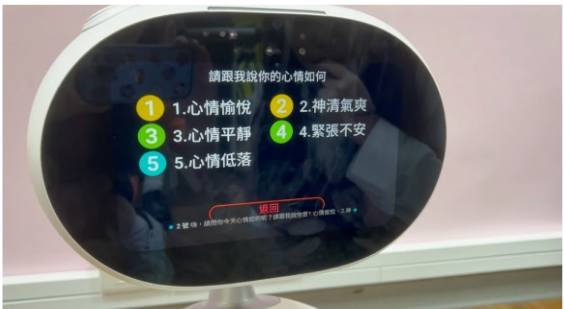

Fig. 7. Social interaction function: mood lift.

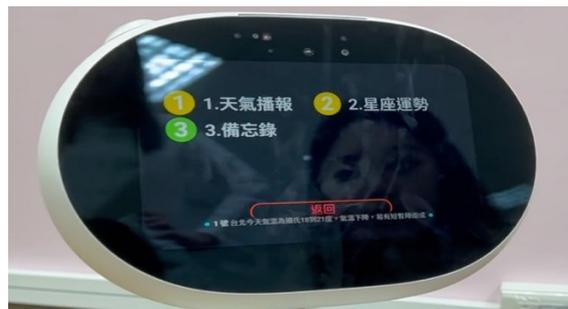

Fig. 9. Social interaction function: today highlight.

In mood lift function, the *Eye Care You* asks questions to visually impaired users, and BSRS (Brief Symptom Rating Scale) is adopted in the question set [17]. The *Eye Care You* asks mood status of visually impaired users, and plays music, video content, joke or talk shows according to their replies (Figure 7). The function is helpful for improving social interactions between visually impaired users and the *Eye Care You*.

In greeting guest function, if there has visitors of visually impaired users, the *Eye Care You* can get to the door and greets to the guest with the word "Welcome" via voice control (Figure 8). The function is useful for the inconveniences of visually impaired users.

### 3.2.2. Today highlight

The DDE (Dialogue Development Environment) editor is adopted for designing contextual scripted conversations, and provides audial content to visually impaired users from the *Eye Care You*. In today highlight function, the content consists of weather forecast, daily horoscopes and daily reminder (Figure 9). Weather forecast provides a weekly weather report of the city where visually impaired users live, and reminds them with the sentences "Today is windy, remember to wear a jacket," or "Today is rainy, remember to bring an umbrella." Daily horoscope includes horoscope information of all signs, and the daily lucky color and lucky number, or what might happen in the future also add fun to their daily lives. In addition, daily reminder is a piratical function for reminding daily schedule of visually impaired users. All of the social interaction functions are triggered via voice control for visually impaired users.

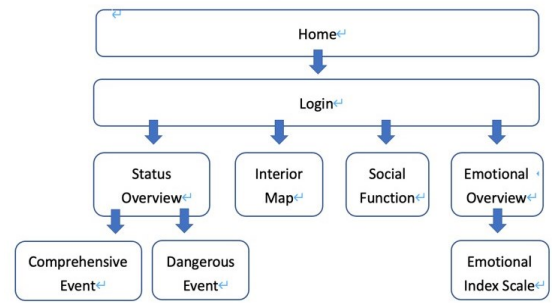

Fig. 10. Sitemap of the website.

### 3.3. Website development

The website is designed for marketing of the *Eye Care You* and also provides a platform to check states of visually impaired users for their caregivers. It was developed by PHP language and SQL database. The website contains four main functions on the home page: status history, indoor map, social function and emotion index (Figure 10). In status history page, it shows photo record of danger events and normal events with time and date information (Figure 11). The record photos are classified by the managers manually. From the record on the website, it helps caregivers to perceive the need of visually impaired users in home environment. Indoor map page shows the scanning result of the home environment with specific marks adding from the caregivers of visually impaired users (Figure 12). Social function page introduces four main social interaction functions of the *Eye Care You* (Figure 13). Emotion index page shows the emotion index of visually impaired users and the correlated videos (Figure 14). The information also helps their caregivers to understand the condition of visually impaired users. Hence, the website provides product information for potential customers and useful information for caregivers of visually impaired users.

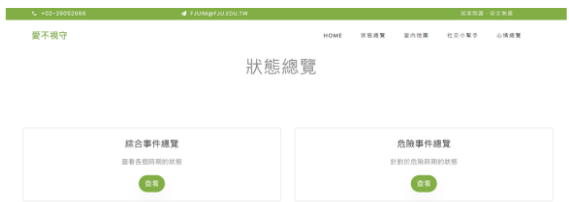

Fig. 11. Status history page of the website.

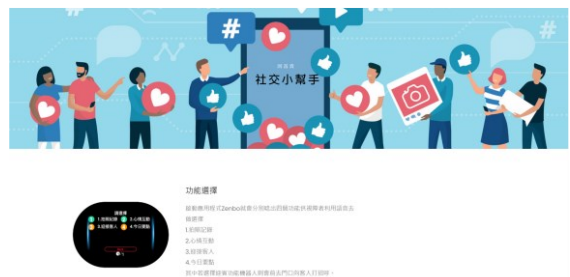

Fig. 13. Social function page of the website.

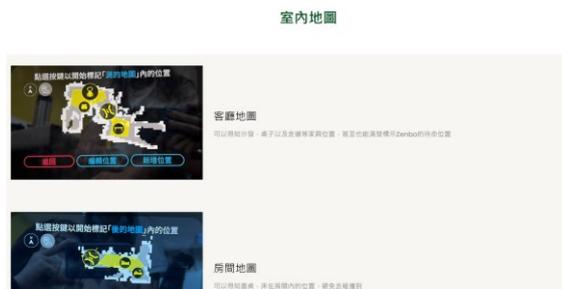

Fig. 12. Indoor map page of the website.

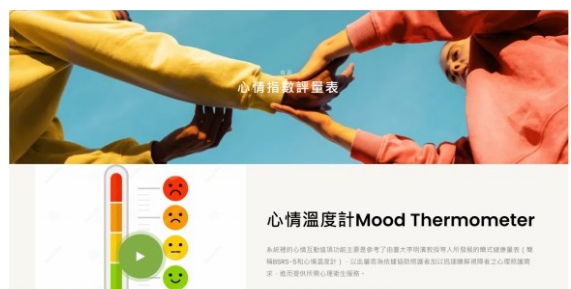

Fig. 14. Emotion index page of the website.

## 4. CONCLUSION

In the study, the *Eye Care You* was provided for improving daily lives of visually impaired users, both considering physical and mental aspects. The *Eye Care You* is robust in the preliminary experiment using in a simulated home scenario. In the future work, it also provides opportunities in the research filed, for example detection of levels of danger events through object recognition technology. Besides, the *Eye Care You* is designed for home scenario currently, and it also can be guided in public space, such as libraries or exhibition halls for visually impaired users in the future.

## REFERENCES

[1] A. Brunes, M. B. Hansen and T. Heir, "Loneliness among Adults with Visual Impairment: Prevalence, Associated Factors, and Relationship to Life Satisfaction," *Health and Quality of Life Outcomes*, Vol. 17, No. 24, pp. 1-7, 2019.

[2] L. C. Weng, "The Way from a Little Room to a Big World: The Experiences of Successful Adaptation with Chronic Interpersonal Problems from a College Student with Visual Impairments," *Journal of Pingtung University of Education: Education*, Vol. 41, pp. 001-034, 2013.

[3] C. A. Chi, "A Study of the Elderly Care System of Visual Impairment," *Bulletin of Special Education*, Vol. 20, pp. 147-169, 2000.

[4] M. Langelaan, M. R. D. Boer, R. M. A. V. Nispen, B. Wouters, A. C. Moll and G. H. M. B. V. Rens, "Impact of Visual Impairment on Quality of Life: A Comparison with Quality of Life in the General Population and with Other Chronic Conditions," *Ophthalmic Epidemiol*, Vol. 14, No. 3, pp. 119-126, 2007.

[5] P. Punia, "Development and Standardization of Mental Health Battery for Visually Impaired," *International Journal of Special Education*, Vol. 33, No. 2, pp. 382-396, 2018.

[6] C. H. Wu and K. Kawauchi, "Influence of Age of Losing Sight and Personal Characteristics on Daily Life Issues of People with Adventitious Visual Impairment," *Journal of Special Education*, Vol. 31, pp. 27-51, 2010.

[7] C. R. Barron, M. J. Foxall, K. V. Dollen, P. A. Jones and K. A. Shull, "Marital Status, Social Support, and Loneliness in Visually Impaired Elderly People," *Journal of Advanced Nursing*, Vol. 19, No. 2, pp. 272-280, 1994.

[8] M. Pinquart and J. P. Pfeiffer, "Mental Well-being in Visually Impaired and Unimpaired Individuals: A Meta-analysis," *British Journal of Visual Impairment*, Vol. 29, No. 1, pp. 27-45, 2011.

[9] G. Vuletić, T. Šarlija and T. Benjak, "Quality of Life in Blind and Partially Sighted People," *Journal of Applied Health Sciences*, Vol. 2, No. 2, pp. 101-112, 2016.

[10] H. Imhonde, A. Olubuogu and L. Handayani, "Personality, Social Support, and Quality of Life as Determinants of Coping Behavior among Visually Impaired Individuals," *Journal of Education and Learning*, Vol. 11, No. 1, pp. 1-8, 2017.

[11] R. Neto and N. Fonseca, "Camera Reading for Blind People, "*Procedia Technology*, Vol. 16, pp. 1200-1209, 2014.

[12] M. Avila, K. Wolf, A. Brock and N. Henze, "Remote Assistance for Blind Users in Daily Life: A Survey about Be My Eyes," *in Proceedings of the 9th ACM International Conference on PErvasive Technologies Related to Assistive Environments (PETRA '16)*, pp. 1-2, 2016.

[13] M. Čaić, D. Mahr and G. O. Schröder, "Value of Social Robots in Services: Social Cognition Perspective," J*ournal of Services Marketing,* Vol. 33, No. 4, pp. 463-478, 2019.

[14] S. Khan, "Towards Improving Human-Robot Interaction for Social Robots, Doctoral Dissertation, University of Central Florida, United States, 2015.

[15] A. A. Scoglio, E. D. Reilly, J. A. Gorman and C. E. Drebing, "Use of Social Robots in Mental Health and Well-Being Research: Systematic Review," *The Journal of Medical Internet Research*, Vol. 21, No. 7, pp. 1-14, 2019.

[16] E. Senft, S. Satake and T. Kanda, "Would You Mind Me if I Pass by You? Socially-Appropriate Behaviour for an Omni-based Social Robot in Narrow Environment," *in Proceedings of the 2020 ACM/IEEE International Conference on Human-Robot Interaction (HRI '20)*, 2020.

[17] M. B. Lee, Y. J. Lee, L. L. Yen, M. H. Lin and B. H. Lue, "Reliability and Validity of Using a Brief Psychiatric Symptom Rating Scale in Clinical Practice," *Journal of the Formosan Medical Association*, Vol. 89, No. 12, pp. 1081-1087, 1990.